\documentclass[aps,prl,twocolumn,groupedaddress]{revtex4-1} 

\usepackage{color}
\usepackage{graphicx}
\usepackage{amsmath}
\begin{document}
\title{Squeezed-light interferometry on a cryogenically-cooled micro-mechanical membrane}
\author{L.~Kleybolte}
\affiliation{Institut f\"ur Laserphysik und Zentrum f\"ur Optische Quantentechnologien, Universit\"at Hamburg, Luruper Chaussee 149, 22761 Hamburg, Germany}

\author{P.~Gewecke}
\affiliation{Institut f\"ur Laserphysik und Zentrum f\"ur Optische Quantentechnologien, Universit\"at Hamburg, Luruper Chaussee 149, 22761 Hamburg, Germany}

\author{A.~Sawadsky$^\S$}
\affiliation{Institut f\"ur Laserphysik und Zentrum f\"ur Optische Quantentechnologien, Universit\"at Hamburg, Luruper Chaussee 149, 22761 Hamburg, Germany}

\author{M.~Korobko}
\affiliation{Institut f\"ur Laserphysik und Zentrum f\"ur Optische Quantentechnologien, Universit\"at Hamburg, Luruper Chaussee 149, 22761 Hamburg, Germany}

\author{R.~Schnabel}
\email{roman.schnabel@physnet.uni-hamburg.de\\[1mm]  $^\S$ Present address: ARC Centre of Excellence for Engineered Quantum Systems, School of Mathematics and Physics, University of Queensland, St Lucia, QLD 4072, Australia\\ }
\affiliation{Institut f\"ur Laserphysik und Zentrum f\"ur Optische Quantentechnologien, Universit\"at Hamburg, Luruper Chaussee 149, 22761 Hamburg, Germany}

\date{\today}

\begin{abstract}
Squeezed states of light reduce the signal-normalized photon counting noise of measurements without increasing the light power and enable fundamental research on quantum entanglement in hybrid systems of light and matter. Furthermore, the completion of squeezed states with cryo-cooling has high potential. First, measurement sensitivities are usually limited by quantum noise \emph{and} thermal noise. Second, squeezed states allow for reducing the heat load on cooled devices without losing measurement precision.
Here, we demonstrate squeezed-light position sensing of a cryo-cooled micro-mechanical membrane. The sensing precision is improved by up to 4.8\,dB below photon counting noise, limited by optical loss in two Faraday rotators, at a membrane temperature of about 20\,K, limited by our cryo-cooler. We prove that realising a high interference contrast in a cryogenic Michelson interferometer is feasible. Our setup is the first conceptual demonstration towards the envisioned European gravitational-wave detector, the `Einstein Telescope', which is planned to use squeezed states of light together with cryo-cooling of its mirror test masses.
\end{abstract}

\maketitle

\section{Introduction}\vspace{-2mm}

Since April 2019, gravitational-wave (GW) events are routinely measured by LIGO and Virgo with signal-to-noise-ratios improved by squeezed states of light on a weekly basis \cite{Tse2019,Acernese2019,Schnabel2020}. The squeezed-light technique \cite{Yuen1976,Caves1981,Schnabel2010,Schnabel2017} allows for a simultaneous reduction of quantum measurement noise (photon counting noise or simply `shot noise') and quantum back-action noise (photon radiation pressure noise) \cite{Jaekel1990}.
The strongest motivation for using squeezed light is the reduction of the signal-normalized photon counting noise since so far it constitutes the dominating quantum noise and the light power cannot be further increased easily in current GW detectors. This issue was also the motivation for implementing squeezed light in the GW detector GEO\,600 in 2010 \cite{LSC2011,Grote2013}.
Very recently, one of the LIGO GW detectors was used to demonstrate the creation of quantum correlations between the differential positions and momenta of the four mirror test masses ($m = 40$\,kg) and the optical field strengths of circulated light ($P = 200$\,kW) at room temperature \cite{Yu2020}.
Future GW observatories, such as the European `Einstein Telescope' \cite{Punturo2010} will employ externally-produced squeezed states to reduce the photon counting noise at audio-band signal frequencies, to reduce photon radiation pressure noise at sub-audio-band signal frequencies, and to employ quantum correlations between optical and mechanical degrees of freedom at frequencies in between. In addition it will utilize cryogenic cooling of the mirror test masses.

In the past, squeezed states of light were ponderomotively produced \cite{Braginsky1967} inside opto-mechanical cavity setups with squeeze factors between 0.06\,dB and 1.7\,dB  at low \cite{Brooks2012,Safavi-Naeini2013,Purdy2013b,Chen2020a} as well as room \cite{Aggarwal2020} temperatures.
Externally-produced squeezed states of light were exploited to improve a optomechanical magnetometer by 2.2\,dB \cite{Li2018} and to suppress quantum back-action in \cite{Yap2020a}, both at room temperature.
So far, there has not been any experiment that demonstrated a laser interferometer with cryo-cooled mirrors and sensitivity enhancement via externally-produced squeezed-light injection.
This confronts the fact that the envisioned `Einstein-Telescope' (ET) as well as `LIGO Voyager'\,\cite{Adhikari2020} build on the combination of these technologies.
The ET pathfinder project is already being set up at the university of Maastricht, and the GW observatory KAGRA has started operation with cryo-cooled mirrors, however, without squeezed light.

Here, we demonstrate the first squeezed-light-enhanced displacement sensing of a cryogenically cooled micro-mechanical SiN membrane at $T = 20$\,K inside a Michelson-Sagnac interferometer.
The membrane had an optical reflectivity of just 19\%, however, the transmitted light did not constitute optical loss, due to the Michelson-Sagnac topology.
The squeezed light enhancement was up to 4.8\,dB around the membrane's fundamental resonance at about 400\,kHz, and equally high over a band of many MHz since no cavity was involved. A Michelson fringe contrast of 98\% was achieved at low temperature.

\section{Experimental setup}
\begin{figure}[]
 \center
     \vspace{0mm}
     \includegraphics[width=8.0cm]{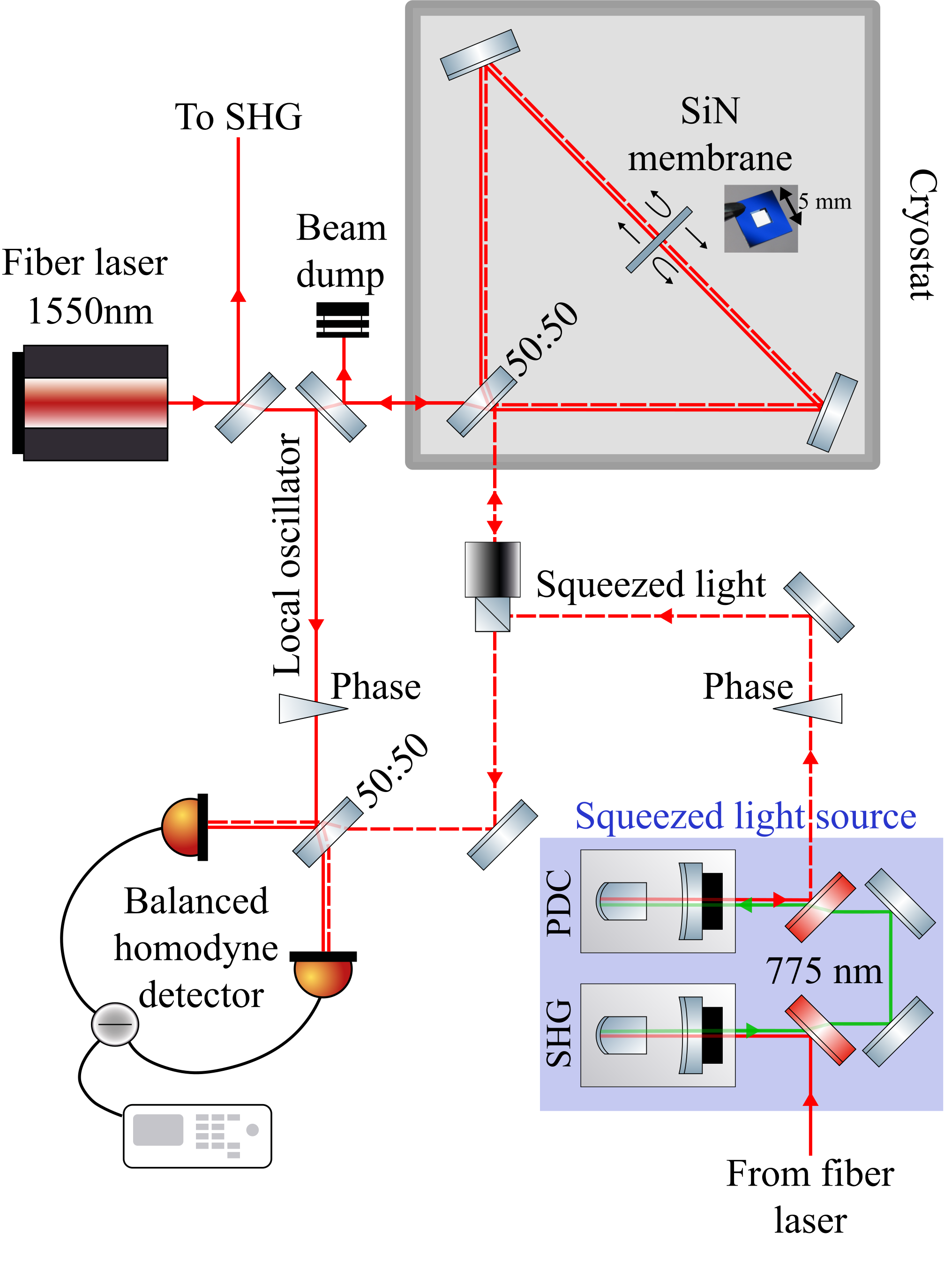}
     \vspace{0mm}
    \caption{Schematic of the experimental setup -- A SiN membrane is part of Michelson-Sagnac interferometer whose solid spacer is attached to a cryo cooler. Similar to the Michelson interferometers used for the observation of gravitational waves, a quasi-monochromatic beam of light carrying coherent states is matched to the overlapping Michelson and Sagnac modes through one port and a squeezed vacuum field is matched to the same modes through the other port. SHG: second harmonic generation; PDC: (cavity enhanced) parametric down-conversion.
    }
    \label{fig:1}
    \vspace{0mm}
\end{figure}

\vspace{-2mm}
We performed an interferometric measurement of the thermally excited motion of a micro-mechanical membrane at a laser wavelength of 1550\,nm.
The light was provided by a commercial fibre laser (\emph{NKT}, model Koheras Boostik) and partially used to produce squeezed states of light in a home-built device.
The membrane acted as a mirror in a Michelson-Sagnac interferometer \cite{Yamamoto2010a}, see Fig.\,\ref{fig:1} for a schematic drawing of the setup.
The displacement of the membrane created a phase modulation on the light that reflected off the membrane.
This reflected light formed the Michelson mode, and the signal was always anti-symmetric in two arms of the interferometer.
The interferometer was tuned as close to a dark fringe of the Michelson mode as possible, such that the membrane signal was fully out-coupled at the signal (dark) port. The Sagnac mode was very close to a dark fringe because the beam splitter was well balanced.
Remaining offsets from the individual-mode dark fringes were set to destructively interfere \cite{Friedrich2011}.
As a result, the DC output power in the signal port of the interferometer was below $1\,\mu$W. This point of operation is in close analogy to current GW observatories, which measure the differential arm length change very close to a dark output port. It is in full analogy to future GW observatories, which need to be operated exactly at dark port to enable balanced homodyne detection for making use of quantum correlations over a broad band of signal frequencies.
\begin{figure}[]
 \center
     \vspace{0mm}
     \includegraphics[width=8.6cm]{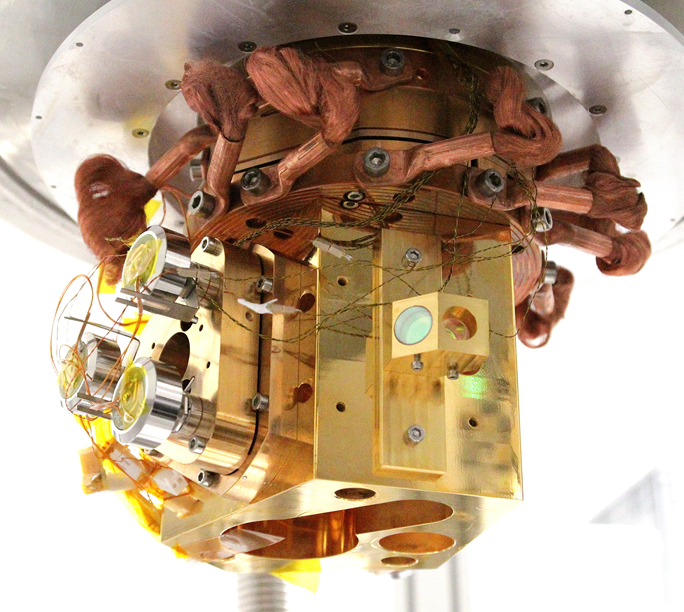}
     \vspace{-3mm}
    \caption{Photograph of the Michelson-Sagnac interferometer -- It contains the SiN membrane and was cooled to about 20\,K by a Gifford-McMahon cryo cooler. Beam splitter and two adjustable steering mirrors are attached to a gold-coated Invar spacer. Membrane position along the optical axis as well as pitch and yaw angles are also adjustable (not visible here).}
    \label{fig:2}
\end{figure}
At this tuning point of an interferometer with equally long arms, the classical frequency and intensity noise of the laser is fully canceled if the alignment is perfect.
The noise is then fully defined by the quantum uncertainty of the vacuum field that enters the interferometer at the signal output port \cite{Caves1981}.
This vacuum field propagates in the interferometer and produces photon radiation pressure noise if the overall light power is high and the mirror masses low. For an interferometer operated at a dark fringe, it finally exits through the signal port together with the signal.
When the signal is measured with a balanced homodyne detector (BHD), which overlaps the signal field with a strong local oscillator (LO), the quantum uncertainty results in photon shot noise in the measurement record.
It constitutes a fundamental limitation of the interferometer sensitivity, overcoming which requires the employment of quantum correlations.

Photon shot noise at a given light power can be overcome using quantum squeezed light.
For that, the vacuum uncertainty need to be squeezed before entering the interferometer.
This squeezing process suppresses the uncertainty in one quadrature of the optical field at a price of increasing it in the orthogonal field quadrature.
The squeezed quadrature is then selected to be in phase with the signal quadrature and the recorded quadrature at the BHD.
Then the shot noise is suppressed without any reduction of the signal strength.

The membrane used in our experiment was translucent, with a reflectivity of about 19\% at 1550\,nm.
When used as an end mirror in one arm of a simple Michelson interferometer, optical loss on signal and light power as well as decoherence on the squeezed states would be intolerable.
In contrast, the Michelson-Sagnac topology \cite{Yamamoto2010}, retains all light, having the capability of zero decoherence on the squeezed states.

The interferometer sensitivity to the displacement of the membrane motion is described by a spectrum $S_x(\Omega)$ versus Fourier frequency of the differential arm length change $\Omega$. It corresponds to the power spectral density of the amplitude quadrature of the light detected on the signal port:
\begin{multline}\label{eq:sens}
  S_{\rm out}(\Omega) = \frac{e^{-2r}\eta}{2}\left(1+t_m + r_m \cos 4 \Delta \right) + r_m \eta \sin^22\Delta \\
  + S_{x}(\Omega)\frac{16\pi r_m P_{\rm in}\eta}{\hbar \lambda_0 c}\cos^2\Delta + 1-\eta,
\end{multline}
where we introduce the squeeze factor $e^{-2r}$, total detection efficiency $\eta$ (including propagation efficiency, mode match between the LO and the signal fields, and quantum efficiency of photodiodes), membrane's power transmissivity and reflectivity $t_m, r_m$, light power $P_{\rm in}$ on the input of the interferometer, the relative phase between the two arms $\Delta$, the laser wavelength $\lambda_0$ and the speed of light $c$.
For the dark port condition, $\Delta=0$, the output signal can be normalized to the membrane motion:
\begin{equation}
  \tilde{S}_x(\Omega) = S_{x}(\Omega) + e^{-2r}\frac{\hbar \lambda_0 c}{16\pi r_m P_{\rm in}},
\end{equation}
where the second term constitutes the quantum noise, here being suppressed by the squeeze factor $e^{-2r}$.

The vibrational mode of our membrane was dominantly driven by thermal Brownian fluctuations coupling from the environmental heat bath.
From the fluctuation-dissipation theorem\,\cite{Callen1952}, the motion of the membrane at the bath temperature T reads:
\begin{equation}\label{eq:fdt}
  S_{x}(\Omega) = \frac{4k_b T\gamma_m}{m_{\rm eff}} \frac{1}{(\omega_m^2-\Omega^2)^2 + \gamma_m^2 \Omega^2},
\end{equation}
where $T$ is the bath temperature, $\omega_m$ is the mechanical resonance frequency, $\gamma_m$ is the mechanical linewidth, $m_{\rm eff}$ is the effective mass of the vibrational mode and $k_b$ is the Boltzmann constant.
\begin{figure}[h!!]
    \includegraphics[width=8.0cm]{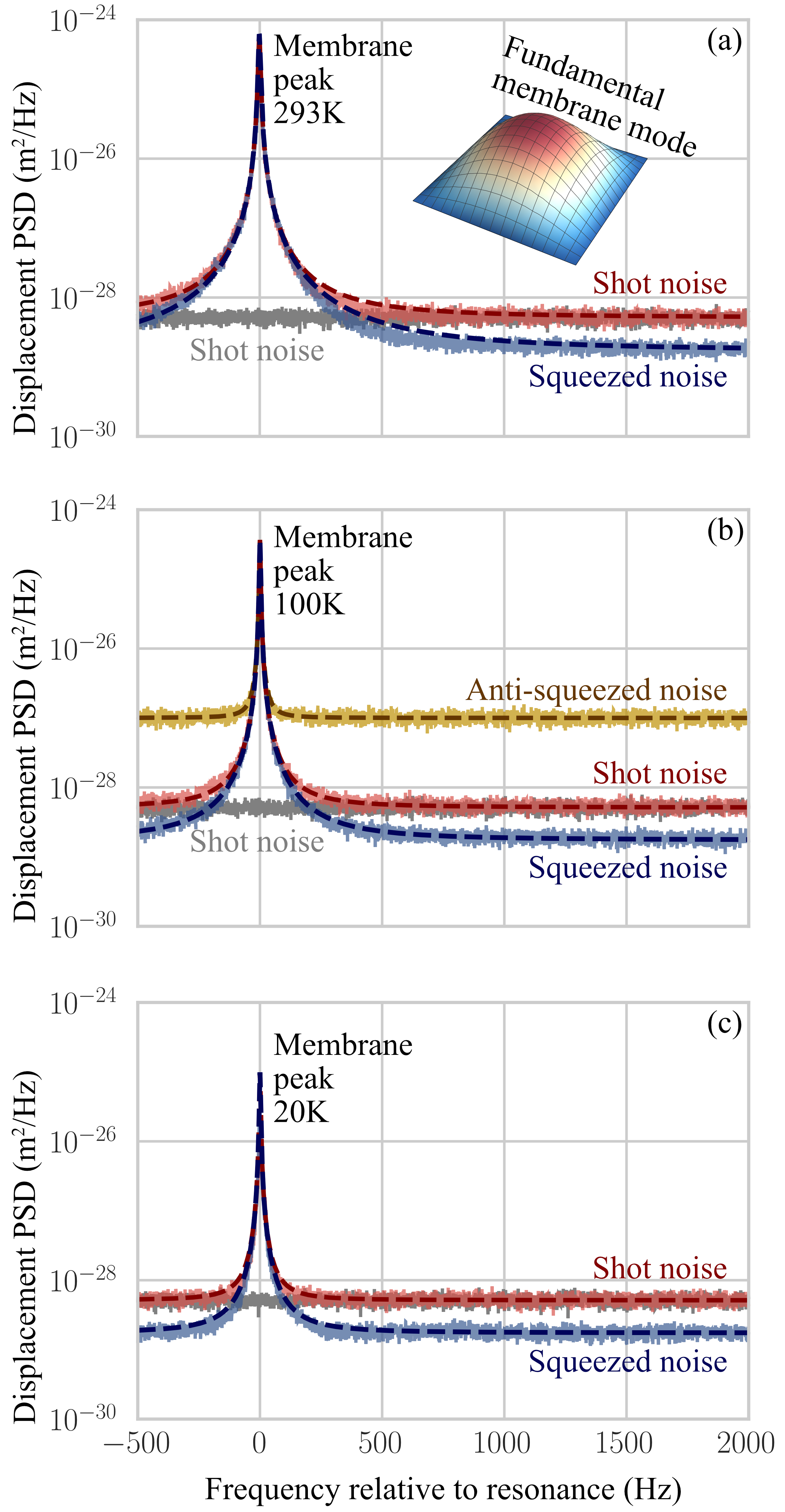}
    \caption{Measured spectra of the same fundamental mode of the membrane oscillation at about 400\,kHz at (a) room temperature, (b) about 100\,K and (c) about 20\,K. We were able to maintain the high interferometer fringe contrast of $>98\%$ for the Michelson mode during the cool-down, which is proven by a constant squeeze factor of about ($4.8 \pm 0.1$)\,dB.
    In (b), we show additionally the anti-squeezed noise, increased by $12.9\pm0.1$\,dB relative to the shot noise.
    The measurements of squeezing and anti-squeezing allow to compute the total optical loss of $\approx31\%$.
    }
    \label{fig:3}
\end{figure}
Following the concepts for future gravitational-wave detectors \cite{Punturo2010},
we cryogenically cooled the membrane down to $\sim 20$\,K to reduce the bath temperature.
In order to maintain the interferometer alignment, we designed a quasi-monolithic interferometer spacer, see Fig.\ref{fig:2}.
The spacer was made out of low-expansion material (Invar) in order to maintain the alignment during the cool-down.
It was coated with gold for shielding against the radiative heating by the warmer environment.
This allowed to reach low temperatures and maintain it when the cryo cooler was turned off during the measurement time, which was necessary for excluding the effect of low-frequency vibrations of the cryogenic pumps.
At low temperatures, we were able to adjust the interferometer alignment using cryogenic positioners from \emph{JPE}.

Squeezed light was produced by type\,0 degenerate cavity-enhanced parametric down-conversion \cite{Wu1986,Vahlbruch2008,Schnabel2017} in periodically-poled KTP \cite{Mehmet2009} from \emph{Raicol}. On our squeezed vacuum source, we observed squeeze factors above 10\,dB at Fourier frequencies of several MHz and up to 8.7\,dB around the fundamental resonance frequency of the membrane at 400\,kHz, using a home-built BHD with reduced electronic noise but slightly less quantum efficiency.
The power spectral density of the photo-electric voltage was analysed by a spectrum analyser.

\section{Results}\vspace{-2mm}
We performed three series of measurements, at room temperature, at about 100\,K and at about 20\,K, all with and without squeezed light injection (Fig.\,\ref{fig:3}).
The peak excitation of the membrane was reduced as we cryogenically cooled the membrane, consistently with the theoretical expectation and Eq.\,\eqref{eq:fdt} for the thermally excited membrane.
The quality factor of the membrane remain at values around $\,10^5$, only slightly varying at different temperatures, and the resonance frequency shifted by a few kHz.
Our setup did not contain any optical cavity, which might have caused optical cooling or heating \cite{Aspelmeyer2014,Sawadsky2015}.
This allowed us to conclude that the membrane was thermally excited, and no other noise significantly contributed to its motion.

We were able to maintain high interferometer fringe contrast of $>98\%$ for the Michelson mode in combination with the Sagnac mode during the cool-down.
The high contrast allowed us to approach the dark signal port condition, and thus reduce the decoherence of the squeezed state of light due to coupling of vacuum field through the imbalance in the interferometer. At all temperatures, we measured about 4.8\,dB of nonclassical quantum noise reduction from injected squeezed vacuum states.
Thanks to the stable interferometer alignment, we were able to retain this level of squeezing at cryogenic temperatures.

Without the interferometer and without transmission through Faraday rotators used for coupling to the interferometer, we observed 8.7\,dB of squeezing from our source at Fourier frequencies around 400\,kHz. Generally, a reduction in the squeeze factor is associated to a combination of higher optical loss, higher technical noise and higher phase noise. In our experiment, the reduction of the squeeze factor was dominated by additional loss due to three transmissions through Faraday isolators ($\sim 12$\%).
Unlike in the idealized case described above, our the interferometer was operated slightly off the perfect dark port condition. The offset produced a small level of carrier light that was used by a servo loop for stabilizing the BHD to the optimal readout quadrature.
This resulted in technical laser power noise with a magnitude not far below the squeezed power spectral density.
Increasing the observed nonclassical shot noise suppression would require low-loss Faraday rotators as they are used in GW observatories, a stabilization of the laser light power in the Fourier frequency band of interest, and stabilizing the BHD readout quadrature by the `coherent control' technique \cite{Vahlbruch2006,Chelkowski2007}, which is used in all current GW observatories.
The absolute sensitivity of our interferometer was also a function of the light power in the arms according to Eq.\,\eqref{eq:sens}.
Here, we used a rather low power in the coherent light to keep the influence of laser power fluctuations low. With stabilized light power and an operation at dark port enabled by the `coherent control' technique, higher light powers would be possible.

\section{Conclusion}\vspace{-2mm}
Any shot-noise-limited laser-interferometric measurement yields a higher sensitivity if optical loss is reduced.
In squeezed-light-enhanced measurements, this effect is the more pronounced the larger the external squeeze factor of the injected light is.
In the setup reported here, the membrane transmitted more than 80\% of the light. To recapture this light, the membrane was put into an Michelson-Sagnac interferometer, whose optical axes of Michelson mode (measurement mode) and Sagnac mode (recapture mode) coincided.
Different from a simple Michelson interferometer, an ideal Michelson-Sagnac interferometer requires a balanced beam splitter, balanced arm lengths, and a beam waist positioned on the translucent membrane. For our cryo-cooled experiment, these additional requirements had to be maintained during cool-down from room temperature to about 20\,K.
We achieved the same modematching of the squeezed-light enhanced arrangement and the same nonclassical improvement of about 4.8\,dB below photon shot noise at 293\,K, 100\,K and 20\,K. This value was limited by optical loss and power fluctuations of the bright light used. The latter can be reduced by stabilizing the laser's output power in the frequency band of interest. The main source of optical loss was located outside the interferometer in terms of two Faraday rotators through which the light had to pass. With  respective reduction of this loss, e.g. by using low-loss devices as installed in current GW detectors the overall loss would be below 10\%, which meets the requirements for the Einstein Telescope.\\

Our result presents the first quantum-enhanced cryogenic interferometer among the upcoming prototypes of future gravitational-wave detectors.
At the same time, it demonstrates the possibility to maintain low decoherence in quantum optomechanical sensors, opening the path towards the integration of the most advanced quantum-optical technologies with the recent achievements of quantum optomechanics.\\

\begin{acknowledgments}
This work was funded by the Deutsche Forschungsgemeinschaft (DFG, German Research Foundation) -- SCHN 757/6-1, and supported by
the DFG under Germany's Excellence Strategy EXC 2121 `Quantum Universe' -- 390833306.
LK and PG were partly supported by the Helmholtz Association through project ExNet-0003.
\end{acknowledgments}

\bibliography{library_RS}

\end{document}